\documentclass{WileyMSP-template}

\usepackage{subfigure}
\usepackage{caption}
\usepackage{array}
\usepackage{tabularx}
\usepackage{multirow}
\usepackage{graphicx,epsfig,epstopdf}
\usepackage{amsmath}
\usepackage{color}
\usepackage{upgreek}
\usepackage{tabularx}
\usepackage[table]{xcolor}
\newcommand{\head}[1]{%
   \textcolor{white}{\textbf{#1}}}

\def\e{\begin{equation}}
\def\f{\end{equation}}
\def\=#1{\overline{\overline #1}}
\def\Re{{\rm Re\mit}}
\def\Im{{\rm Im\mit}}

\def\_#1{{\bf #1\mit}}

\def\l#1{\label{eq:#1}}
\def\r#1{(\ref{eq:#1})}

\newcolumntype{Y}{>{\centering\arraybackslash}X}

\begin{document}


\title{Three-dimensional random dielectric colloid metamaterial\\ with giant isotropic optical activity}

\maketitle




\author{Viktar~S.~Asadchy}
\author{Cheng~Guo}
\author{Ihar~A.~Faniayeu}
\author{Shanhui~Fan}


\dedication{}

\begin{affiliations}
Dr.~V.S. Asadchy,  Prof.~S. Fan\\
Ginzton Laboratory and Department of Electrical Engineering, Stanford University, Stanford, California 94305, USA\\
Email Address: asadchy@stanford.edu; shanhui@stanford.edu

C. Guo\\
Department of Applied Physics, Stanford University, Stanford, California 94305, USA

Dr.~I.A.~Faniayeu\\
Department of Physics, University of Gothenburg,
 Gothenburg 412~96, Sweden

\end{affiliations}



\begin{abstract}

Motivated by the  theoretical observation that isotropic chirality can exist even in completely random systems, we design a dielectric metamaterial consisting of a random colloid of meta-atoms, which exhibits unprecedentedly high isotropic optical activity. Each meta-atom is composed of a helically arranged cluster of silicon nanospheres. Such clusters can be fabricated by large-scale DNA self-assembly techniques.  It is  demonstrated that the use of a high concentration of the meta-atoms in the colloid provides significant suppressions of incoherent scattering losses. As a result, the proposed system shows three orders of magnitude improvement of isotropic optical activity as compared with the previous metamaterial designs. This work highlights the significant potential of completely random system, which are  commonly produced in colloidal sciences, for applications as metamaterials towards novel photonic effects and devices. 

\end{abstract}


\section{Introduction}
The development of metamaterials opens numerous new opportunities for manipulating electromagnetic waves. Most metamaterial structures consist of regular array of meta-atoms~\cite{soukoulis_past_2011}. As an alternative, one may wonder whether or how one can achieve non-trivial metamaterial functionalities in complete random or disordered systems. This is important from a fundamental perspective in the context of recent efforts in exploring the effects of disorders in photonic systems~\cite{albooyeh_resonant_2014,titum_disorder-induced_2015,bandres_topological_2016,agarwala_topological_2017,xiao_photonic_2017,liu_disorder-immune_2019}. It is also of significance from a practical point of view, since many large-scale fabrication techniques, such as self-assembly techniques, readily produce random array of meta-atoms~\cite{xia_self-assembly_2001,kim_self-assembled_2011,arpin_three-dimensional_2013,cecconello_chiroplasmonic_2017,liu_dna-assembled_2018,kuzyk_dna_2018}.

For a linear metamaterial consisting of a completely random array of meta-atoms, after homogenization, its electromagnetic properties can be described by an effective medium with the following   constitutive relations~\cite[p.~28]{serdyukov_electromagnetics_2001}:
\begin{equation}
\begin{array}{c}
    \_D=\varepsilon \cdot \_E + \xi \cdot \_H, \\
    \_B=\mu \cdot \_H + \zeta \cdot \_E.
    \end{array}
    \label{first}
\end{equation}
Here $\_D$, $\_B$, $\_E$, and $\_H$ are the electric and magnetic displacement vectors and fields, $\varepsilon$ and $\mu$ are the permittivity and permeability tensors, respectively, and  $\xi$ and $\zeta$ are the bianisotropic tensors.  If one further assumes that the effective medium is lossless, then $\xi = \zeta^\dag$~\cite[\textsection~2.7]{serdyukov_electromagnetics_2001}.
Since the array is completely random, the effective medium must be rotationally invariant. Consequently, $\varepsilon$, $\mu$ and $\xi$   must all be scalar.  In the lossless case,  the imaginary part of $\xi$ describes effect of isotropic chirality (natural optical activity), which is a reciprocal effect, while the real part of $\xi$ describes the effect of isotropic    magneto-electric coupling (alternatively called Tellegen effect), which is a nonreciprocal effect.   Therefore, one should be able to observe isotropic chirality or isotropic magneto-electric coupling  even in a system that is completely random.

Motivated by the theoretical considerations above, in this Letter, we introduce a dielectric metamaterial consisting of a random colloid of specifically designed  silicon nanoclusters in order to achieve giant isotropic chirality. 
We demonstrate that due to the absence of absorption, the concentration of the nanoclusters  in the colloid can be very high, which, in turn, provides a significant suppression of   incoherent scatterings. 
Our metamaterial colloid   exhibits isotropic chirality that is   three orders of magnitude higher than those of previously reported isotropic chiral materials.   Moreover, we demonstrate that similar superiority is achieved in terms of the  figure of merit  given by the ratio of the chirality parameter and the average extinction coefficient (such figure of merit is introduced in analogy to the magneto-optical figure of merit~\cite[\textsection~9.6.5]{zvezdin_modern_1997}).
Since we consider in this contribution only isotropic chirality effect and neglect the nonreciprocal magneto-electric effect, in what follows, we use the conventional electromagnetic notation  according to which scalar parameters $\xi=i \kappa/ c$ and $\zeta= - i \kappa/c$ are expressed in terms of  the so-called chirality parameter $\kappa$~\cite[p.~28]{serdyukov_electromagnetics_2001} ($c$ is the speed of light).  Time harmonic oscillations in the form $e^{-i \omega t}$ are assumed throughout the Letter.

Chirality phenomenon, first observed more than two centuries ago~\cite{arago_memoire_1811},  is of current importance for a diversity of applications such as in  organic and inorganic chemistry,  pharmaceutical engineering, optical communications, displays, spectroscopy, and  sensors~\cite{barron_molecular_2004,schaferling_chiral_2017,hentschel_chiral_2017,caloz_electromagnetic_2019}.
The chiral effects in naturally occurring materials are relatively weak. Therefore, there is significant recent interest in designing metamaterials~\cite{lindman_uber_1920,tinoco_optical_1957,ougier_measurement_1992,guerin_microwave_1994,tretyakov_waves_2003,soukoulis_past_2011,kaschke_optical_2016,wang_optical_2016,fernandezcorbaton_new_2019} 
and colloids~\cite{mastroianni_pyramidal_2009,kuzyk_dna-based_2012,shen_three-dimensional_2013,kaschke_optical_2016,cecconello_chiroplasmonic_2017,kuzyk_dna_2018,liu_dna-assembled_2018}
with  enhanced chiral response. In particular, there is an extensive literature on the chiral properties of metasurfaces~\cite{wu_spectrally_2014,verre_metasurfaces_2017,singh_large_2018,xiao_giant_2018}. The chiral response of a three-dimensional metamaterial, however, can be qualitatively different. Isotropic chiral response, where the same chiral effects occur independently of the directions of light propagation, is a three-dimensional effect that is difficult to achieve with metasurfaces. Such an isotropic chiral response is potentially important for topological photonics~\cite{gao_topological_2015,xiao_hyperbolic_2016} and stereochemistry~\cite{hentschel_chiral_2017,solomon_enantiospecific_2019}. Moreover, isotropic chiral materials  provide an attractive route towards achieving the phenomenon of negative refraction   in bulk~\cite{tretyakov_waves_2003,pendry_chiral_2004}.  
In the optical frequency range, the existing designs of three-dimensional chiral metamaterials rely upon metallic structures with complex geometries~\cite{soukoulis_past_2011,kaschke_optical_2016,wang_optical_2016,fernandezcorbaton_new_2019}. The use of metal introduces significant inherent loss, whereas the use of complex geometries represents  major challenges for three-dimensional nanofabrication. Our proposed chiral metamaterial colloid overcomes all the above-mentioned drawbacks.

\section{Results}
\subsection{Meta-atom design}
In an isotropic chiral material, the right (RCP) and left (LCP) circularly polarized light are  eigenstates for all propagation directions. 
The chirality parameter of a material is defined as   $\kappa = (n_- - n_+)/2$, where $n_+$ and $n_-$ are the refractive indices for right   and left   circularly polarized   light, respectively~\cite{caloz_electromagnetic_2019}. The isotropic chirality results in the optical rotation~$\Delta\theta$ and the circular dichroism~$\eta$  effects for waves propagating over distance~$L$ in the material, which is expressed as $\Delta\theta +i \eta= \omega \kappa L/c$~\cite[\textsection~5.2]{barron_molecular_2004}.   For a metamaterial, the chirality parameter is nonzero  only when the constituent meta-atoms have no mirror symmetry.  
For low-concentration colloids,  the macroscopic chirality parameter   and the properties  of the individual meta-atoms are related by~\cite[\textsection~2.10.1]{serdyukov_electromagnetics_2001}
\e 
\kappa= \frac{N}{3} {\rm tr} \{ \={\alpha}_{\rm em} \},
\l{trace1}
\f
where $N$ is the volume concentration of the constituent meta-atoms  in the colloid and $\={\alpha}_{\rm em}$ is the magnetoelectric polarizability tensor of each meta-atom. This relation is equivalent to the Rosenfeld equation in chemistry literature~\cite[Eq.~5.2.2]{barron_molecular_2004}. For high constituent concentrations, the chirality parameter  is obtained using the   Maxwell Garnett mixing rule~\cite[\textsection~7.2.2]{serdyukov_electromagnetics_2001}.  
Relation~\r{trace1} is rotationally invariant  since the trace of a matrix does not change under  rotations. 

As is seen from~\r{trace1}, when  choosing the constituent meta-atom for our colloid, we need to maximize  the trace of its magnetoelectric polarizability tensor~$\={\alpha}_{\rm em}$. Next, we suggest a simple extraction technique of this trace for an arbitrary  meta-atom. The technique is based on  a previously proposed method based on the  far-field scattering~\cite{asadchy_determining_2014}. 
Consider an arbitrary meta-atom in free space.
\begin{figure}[tb]
\centering
   \includegraphics[width=0.46\columnwidth]{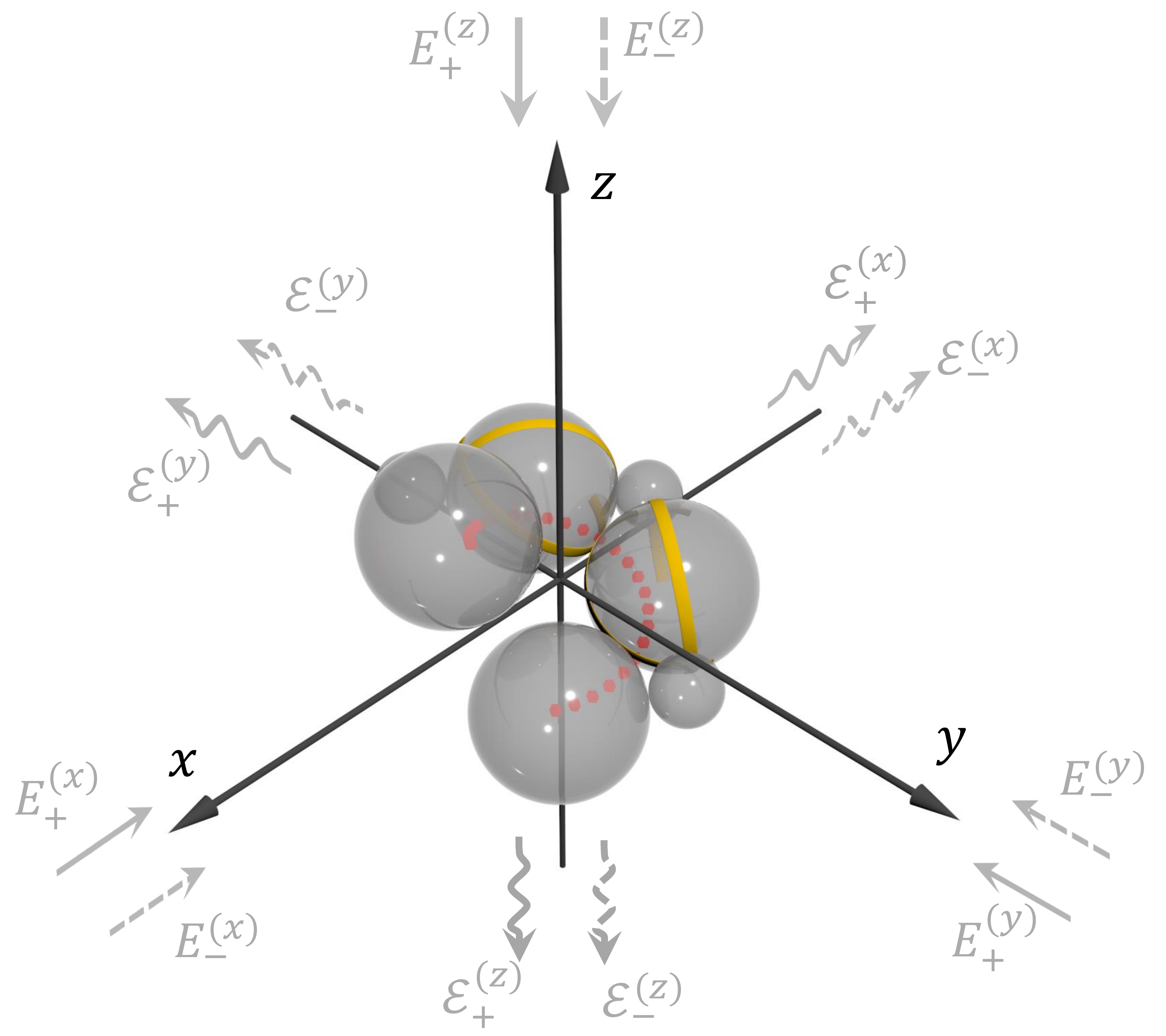}
\caption{  Designer dielectric chiral  meta-atom assembled as a cluster of crystalline silicon nanospheres. The nanospheres are DNA-coated and hybridized by DNA origami belts (shown as yellow strips) to  one another, forming a three-dimensional helical shape (shown by red dots). The straight  and curvy arrows denote the directions of the incident and forward-scattered light, respectively. The plus and minus subscripts stand for RCP and LCP light, respectively. Six measurements of the forward-scattered light with two circular polarizations and along three illumination directions enable  accurate and unambiguous calculation of the intrinsic (isotropic) chiral properties of the meta-atom.  }
\label{fig1}
\end{figure}
We illuminate it alternately with RCP and  LCP   monochromatic light  along the $-x$, $-y$, and $-z$ directions, as shown with gray arrows in Fig.~\ref{fig1}.  The upper and bottom signs in the electric field amplitudes $|\_E_\pm|= \sqrt{2} E_0$   correspond to the right and left circular polarizations, respectively. Next, we can  arrive   at the following expression for the trace of the polarizability tensor~$\={\alpha}_{\rm em}$ (see Supporting Information~\cite[\textsection~1]{suppl}):
\e 
{\rm tr} \{ \={\alpha}_{\rm em} \} = 
\frac{ \pi  r c^2}{\omega^2 E_0 e^{i \omega r/c}} \sum_{m=x,y,z}
\left(  \mathcal{E}_+^{(m)} -  \mathcal{E}_-^{(m)} \right),
\l{trace3}
\f
where $\mathcal{E}_+$ and $\mathcal{E}_-$ are the   RCP  forward-scattered field from the meta-atom when illuminated by RCP light and the LCP    forward-scattered field from the meta-atom when illuminated by LCP light, respectively. Subscript $m$ denotes the direction of the incident light. The scattered light can be measured using full-wave simulations at a distance $r$ from the meta-atom.  
Expression~\r{trace3} allows simple calculation of the trace of  the magnetoelectric tensor  of an unknown meta-atom from six scattering calculations. As one can see, the trace as well as optical activity (according to relation~\r{trace1}) are maximized when the meta-atom exhibits the highest contrast in forward scattering of incident RCP and LCP   light for the three orthogonal directions.  
It is worth mentioning that the derived expression~\r{trace3} to some extent resembles the conventional
dimensionless   circular intensity difference parameter 
defined as $\Delta=(I_+ -I_-)/(I_+ + I_-)$~\cite[p.~162]{barron_molecular_2004}, where $I_+$ and  $ I_-$ are the scattering intensities of the meta-atom excited by the incident RCP and LCP light, respectively. 

Next, we design the  constituent meta-atom   in the form of a cluster of dielectric nanospheres. In order to maximize the interaction between the meta-atoms and incident light, we need to use high-permittivity dielectric. For the chosen operational frequency of 400~THz ($\lambda_0=750$~nm wavelength in vacuum or $\lambda=564$~nm in water where the meta-atoms are dispersed), a good candidate is crystalline silicon~\cite{schinke_uncertainty_2015,baranov_all-dielectric_2017}.   Fabrication of both crystalline and amorphous silicon nanosphere colloids have been demonstrated in the literature~\cite{fenollosa_silicon_2008,li_fabrication_2011,shi_monodisperse_2013}. The nanospheres in the colloid can be coated by DNA origami belts and hybridized to form a cluster with  the desired chiral shape, as was demonstrated recently in~\cite{zion_self-assembled_2017}. Some alternative fabrication techniques for such a colloid metamaterial can be found in~\cite{fu_discrete_2004,yan_self-assembly_2012,ma_electric-fieldinduced_2015,zerrouki_chiral_2008,freire_nanospheres_2012,fan_self-assembled_2010,wang_synthetic_2015}.
We choose the maximum size of the cluster meta-atom $D_{\rm max}=130$~nm so that it remains subwavelength in water at wavelength $\lambda=564$~nm. Such choice, although not unique, was made in order to avoid generation of higher multipoles in the meta-atoms and diffraction effects, as well as to have the possibility of strong suppression of light scattering in the colloid metamaterial, as discussed below. With such a choice of sizes, the designed meta-atom is nonresonant: The first Mie resonance for a quasispherical nanoparticle at the chosen operating  frequency occurs when its  size is $D_{\rm Mie}\approx \lambda_0/ n_{\rm Si}=202$~nm~\cite{kuznetsov_optically_2016}, which is   twice   larger than our chosen size $D_{\rm max}$. 

Using the expression~\r{trace3} and calculating the six scattered far-fields $\mathcal{E}_+$ and $\mathcal{E}_-$ via ANSYS HFSS (a finite element method solver), we have found that  large  value  of the trace of $\={\alpha}_{\rm em}$ tensor  (for the given maximum  size of the occupied volume $\pi D_{\rm max}^3/6$) is achieved for a helically-shaped cluster of nanospheres (helical shape was chosen as it is considered optimal for metalic structures~\cite{fernandez-corbaton_objects_2016,cecconello_chiroplasmonic_2017}). The designed chiral meta-atom is shown in Fig.~\ref{fig1} and consists of four larger nanospheres with diameters~$D=52$~nm and three smaller nanospheres with diameters~$d=20$~nm. Note that for simplifying  fabrication, the   three smaller  nanospheres could be omitted, which would decrease the trace of $\={\alpha}_{\rm em}$ by approximately   half. An imaginary line drawn through the centers of the larger nanospheres forms a helical trajectory with pitch of 58~nm and diameter of 66~nm (the neighbouring nanospheres are touching one another and the centers of the two  outermost spheres are at the ends of the helical trajectory), as shown by the red dotted line in the figure. Each smaller nanosphere touches two  larger ones such that a plane drawn through their centers (one smaller and two larger nanospheres) passes through   the origin of the chosen coordinate system (see Fig.~\ref{fig1}).
Table~\r{tab1} summarizes the traces of the dipolar electric, magnetic, and magnetoelectric polarizability tensors extracted   using \r{trace3} and technique reported in~\cite{asadchy_modular_2019}. 
\begin{table}[tb]
   \centering
   \sffamily
\begin{tabularx}{\columnwidth}{ YYY }  
     \rowcolor{black!75}
  \head{$ {\rm tr} \{ \overline{\overline{\alpha}}_{\rm ee} \}/3 $ ${[\rm m}^3]$} & \head{$ {\rm tr} \{ \overline{\overline{\alpha}}_{\rm mm} \} /3$ ${[\rm m}^3]$} & \head{$ {\rm tr} \{ \overline{\overline{\alpha}}_{\rm em} \} /3$ ${[\rm m}^3]$}  \\
   \rowcolor{black!15}
     $(8.3+i0.3) \cdot 10^{-22} $ & $(5.4+i0.3) \cdot 10^{-23} $       & $(-8.0+i0.2) \cdot 10^{-25} $ \\
  \end{tabularx}
  \caption{ Dipolar polarizabilities of the designer dielectric meta-atom shown in Fig.~\ref{fig1} averaged over all orientations at the operational frequency of 400~THz.  }
  \l{tab1}
\end{table}
As is seen, the largest isotropic response arises from the electric dipole excitation, exceeding by one or several orders of magnitude those of the magnetic dipole excitation and magnetoelectric coupling. Such unbalanced polarization response    indicates that the designed cluster is still far from the theoretical maximum chirality bound which occurs when $ \alpha_{\rm ee} = \alpha_{\rm em} =\alpha_{\rm mm}$~\cite{semchenko_radiation_2004,radi_balanced_2013,fernandez-corbaton_objects_2016}. Nevertheless, this theoretical bound can be achieved only with  strong field localization, like in plasmonic meta-atoms. However, the lossy nature of plasmonic clusters is inconsistent with the goal of the present work.
The imaginary part of the trace of the polarizability takes into account both the scattering of light away from the direction of incidence, which becomes the incoherent scattering loss for the random metamaterial, as well as material absorption loss since the operating frequency is above the band gap of crystalline silicon. For the proposed meta-atom, the   scattering loss dominates over the absorption loss by two orders of magnitude.  
Additionally, we have analyzed a pyramidal cluster of four touching nanospheres (a cluster of three arbitrary nanospheres always has   the inversion symmetry). The nanospheres were chosen with the same radii of 30~nm, while their permittivities were all different to break inversion symmetry: $\varepsilon_1=2$, $\varepsilon_2=6$, $\varepsilon_3=10$, and $\varepsilon_4=13.8$. For this cluster, the absolute values of the polarizability traces are ${\rm tr} \{ \overline{\overline{\alpha}}_{\rm ee} \}/3 =8.2\cdot 10^{-22}~{\rm m}^3$, ${\rm tr} \{ \overline{\overline{\alpha}}_{\rm em} \}/3 =2.0\cdot 10^{-25}~{\rm m}^3$, and ${\rm tr} \{ \overline{\overline{\alpha}}_{\rm mm} \}/3 =5.2\cdot 10^{-23}~{\rm m}^3$. Comparing the obtained value of   the magneto-electric polarizability trace to that  in Table~\r{tab1}, one can see that the pyramidal cluster exhibits four times weaker chiral properties than the helical cluster, which is in agreement with the results in~\cite{cecconello_chiroplasmonic_2017}.

\subsection{Scattering inside a random metamaterial and supercell approximation}


Next, we study optical response of the random distribution of the designed dielectric constituent meta-atoms, as shown in Fig.~\ref{fig1}, dissolved in   water. From \r{trace1}, since the strength of the isotropic chirality parameter increases as the volume concentration $N$ increases, we are interested in the regime with the high volume concentration. As a dimensionless measure of the volume concentration, we define $N_{\rm V}=N V_{\rm circ} $, where $V_{\rm circ}=1.15\cdot 10^{-21}~{\rm m}^3$ is the volume of the sphere circumscribed around the meta-atom shown in Fig.~\ref{fig1}.
For given volume concentration, we analytically calculate the relative permittivity~$\=\varepsilon$, permeability~$\=\mu$, and chirality parameter~$\=\kappa$ of the metamaterial using the Maxwell Garnett mixing rule~\cite[\textsection~7.2.2]{serdyukov_electromagnetics_2001}, whose applicability is not limited to low concentrations. Effective refractive indices for RCP and LCP light propagating in the colloid metamaterial are expressed as $n_\pm= \sqrt{\varepsilon \mu} \mp \kappa$~\cite[\textsection~2.2.1]{lindell_electromagnetic_1994}. In order to calculate the figure of merit, we consider  the   attenuation coefficient~$\beta=2 k_0 \Im(n)$ of the colloid metamaterial which shows how much light is attenuated over propagation distance $L$, i.e. $I(L)=I(0) e^{-\beta L}$. As mentioned above, the primary origin of the attenuation is from incoherent light scattering, which arises from  the irregular  distribution of the constituent meta-atoms in the metamaterial (fluctuations of density of scatterers).
The incoherent scattering occurs even in the situations when  the average distance between the meta-atoms is much smaller than the wavelength  and the medium can be considered as uniform~\cite{sobelman_theory_2002}. 

\begin{figure}[tb]
\centering
	\centering
	\subfigure[]{\includegraphics[width=0.45\linewidth]{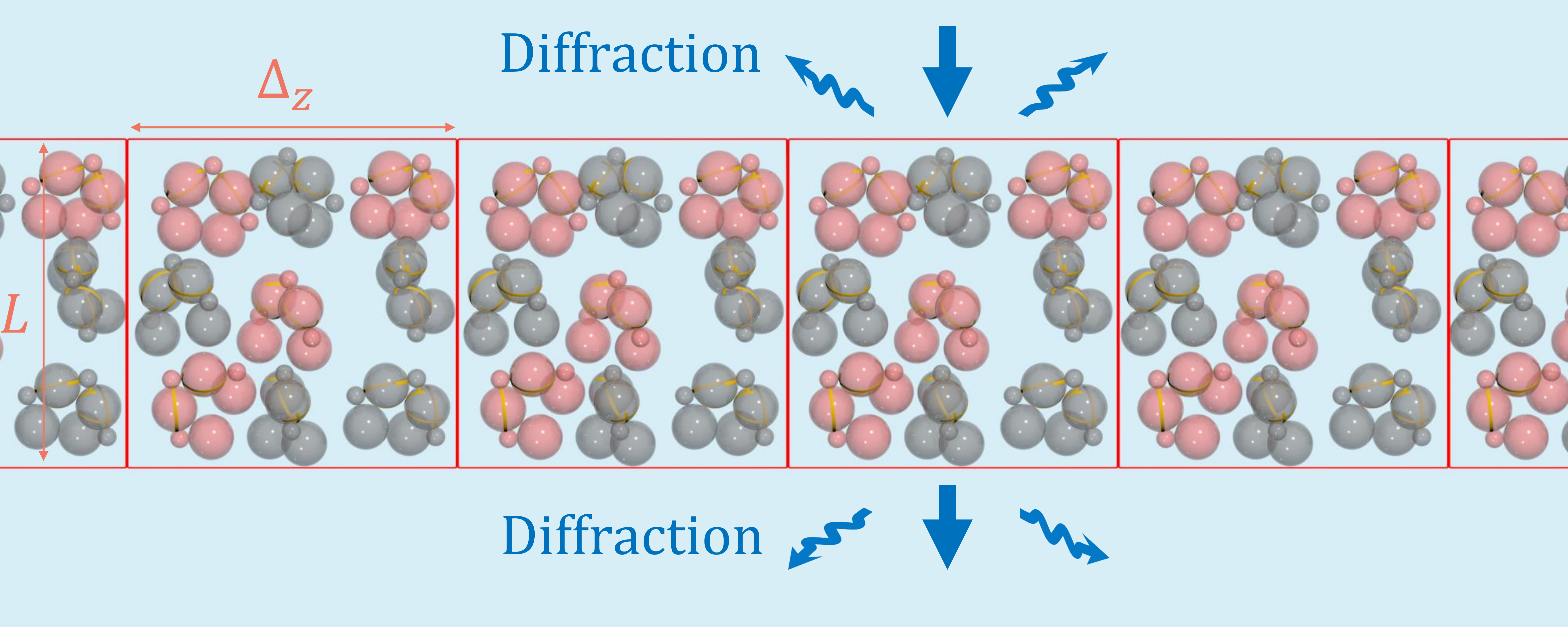} \label{supercell1}}  \\
	\subfigure[]{\includegraphics[width=0.45\linewidth]{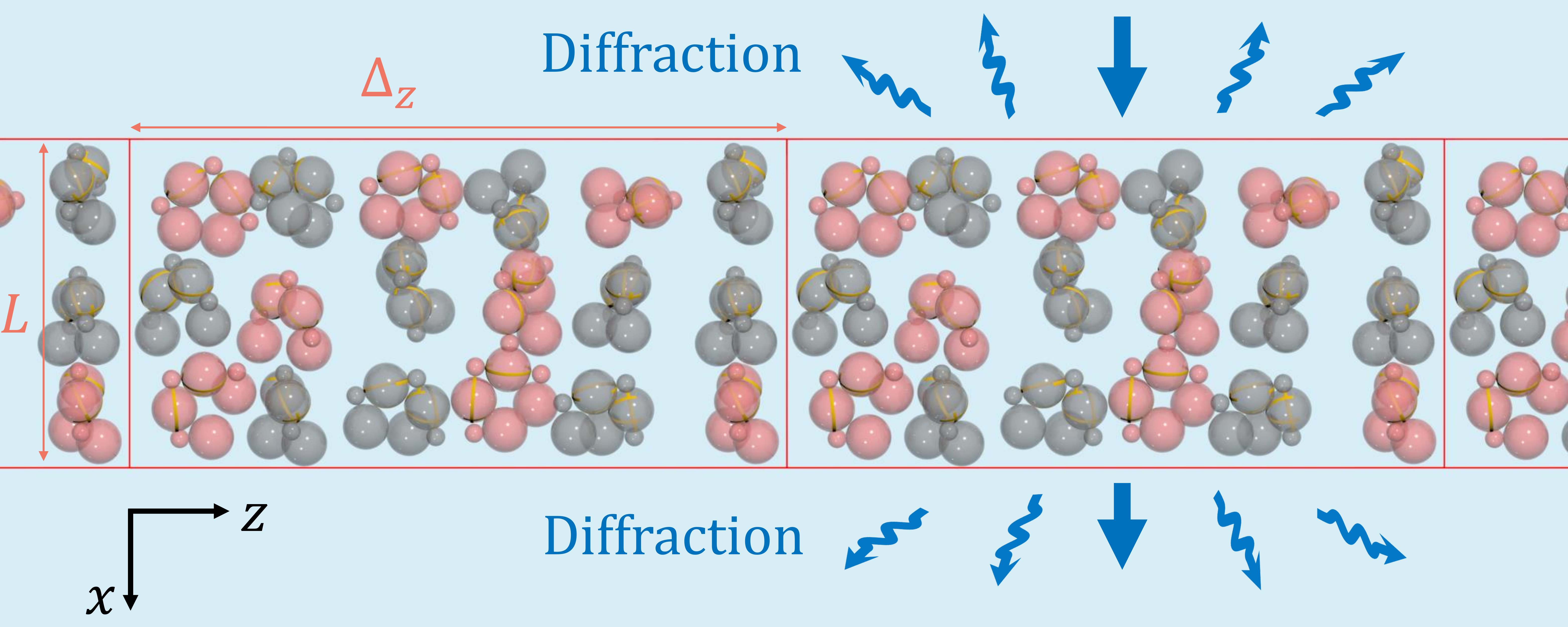}\label{supercell2}}  \\
	\subfigure[]{\includegraphics[width=0.45\linewidth]{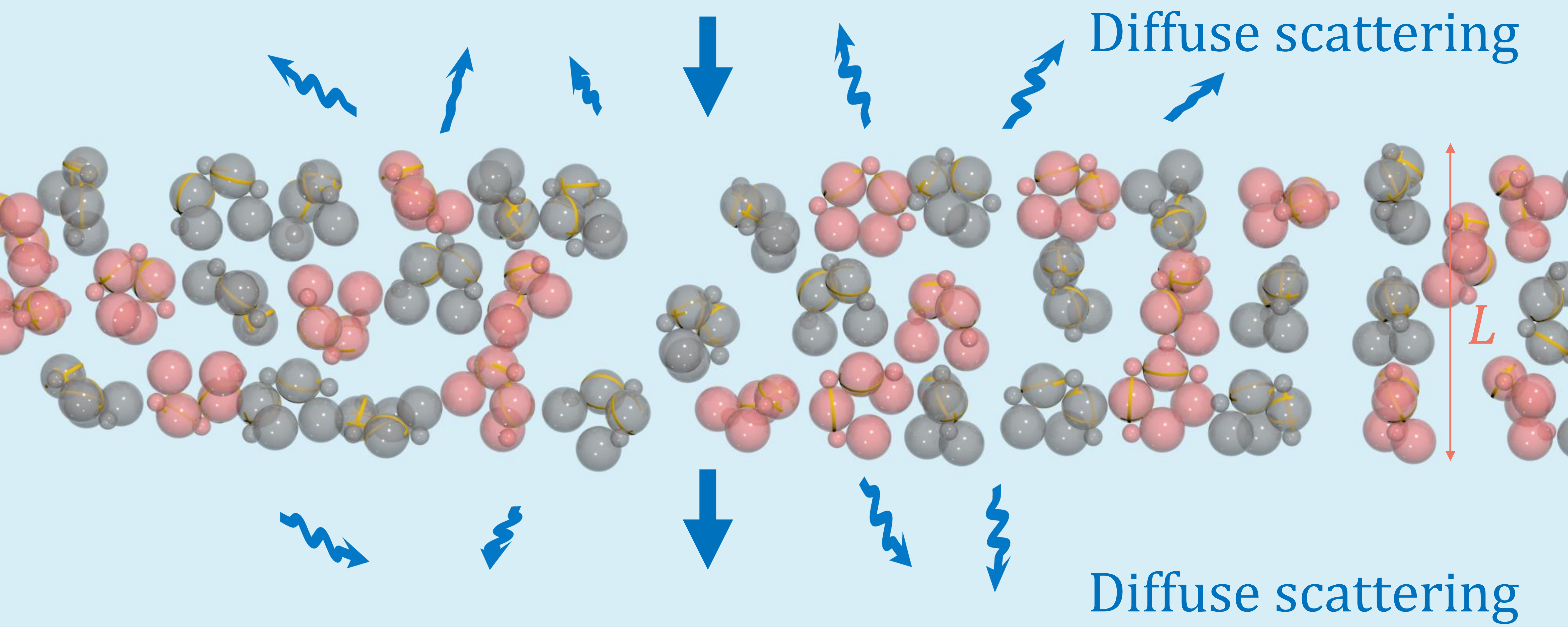}\label{supercell3}}
\caption{  Illustration of the supercell approximation applied for the scattering problem from a random colloid metamaterial. The meta-atoms were painted in two different colors for easier recognition of  the supercell.    (a) A periodic metamaterial slab of thickness $L$ with the supercell (shown as a red box) repeated in the $y$ and $z$ directions. A part of the normally incident light is scattered into four diffraction orders. (b) Same as (a) but with larger supercell and eight diffraction orders. (c) A metamaterial slab which is periodic in the $y$-  but  completely random in the $z$-direction.   In all the subfigures, the    volume fraction $N_V=0.6$ is chosen.  
In   the supercell approximation, diffuse scattering from the random metamaterial in (c)  is studied by considering the scattering in various diffraction orders of the periodic metamaterials in (a) and (b). 
   }
\label{supercell}
\end{figure}

According to the classical independent-scattering formulation of the radiative transfer theory, the total scattering in a colloid is proportional to the volume concentration of its  constituent meta-atom~\cite{twersky_transparency_1975,ishiniaru_attenuation_1982,tsang_effective_1982,tsang_monte_1992,west_comparison_1994}. Under this assumption, the scatterings from all the meta-atoms  are uncorrelated and  $\beta_{\rm uncorr} (N)=2 k_0 \Im(n)= N \sigma$, where $\sigma$ is the extinction cross section of individual meta-atom. 
However, as it was demonstrated in the early works of Twersky et al.~\cite{hawley_comparison_1967,twersky_transparency_1975,twersky_acoustic_1978}, this assumption is accurate only for media with low  volume fraction $N_{\rm V}$ (volume fraction is defined as a product of the volume of a single meta-atom and the volume concentration of the colloid, that is $N_{\rm V}=V_0 N$). In dense colloids, there exist correlations in   positions of the meta-atoms which lead to the modified attenuation coefficient:
\e
\beta_{\rm corr}(N, N_{\rm V}) =N \sigma W(N_{\rm V})= 2 k_0 \Im(n_{\rm corr}),
\l{corr}
\f
where $W(N_{\rm V})=  (1-N_{\rm V})^4/(1+2N_{\rm V})^2$ is the so-called packing factor for spherical-shape constituent meta-atoms~\cite{twersky_transparency_1975} and $\Im(n_{\rm corr})=\Im(n) W(N_{\rm V})$ is the effective refractive index of the colloid metamaterial whose imaginary part is scaled by the packing factor. The factor $W(N_{\rm V})$ monotonically decreases with increasing volume fraction, leading to the fact that attenuation due to incoherent scattering in dense colloid  metamaterials can be several orders of magnitude smaller than that predicted by the independent-scattering theory. Such scattering reduction can be intuitively understood by increased regularity of the colloid due to  the loss of available space for its constituent meta-atoms. Equation~\r{corr}, which was sometimes referred to as the quasicrystalline approximation (QCA),  was subsequently confirmed for dense media   experimentally~\cite{ishiniaru_attenuation_1982,west_comparison_1994} and numerically~\cite{tsang_monte_1992}.

In our study, we exploit the effect of scattering loss reduction to design a  chiral dielectric  colloid metamaterial with high figure of merit, that is  ${\rm FOM}=|\kappa|/\Im(n_{\rm corr})$. This FOM  is introduced in analogy to the magneto-optical FOM, where the latter  is defined as the specific Faraday rotation divided by the attenuation constant~\cite[\textsection~9.6.5]{zvezdin_modern_1997}. 
For our system, the maximum volume fraction of the colloid is approximately $N_{\rm V, max} = 0.63$, which is the measured value  of $N_V$ for   densely packed amorphous solids of identical spheres~\cite{scott_packing_1960}. This volume fraction corresponds to that of the mean of the cubic   and hexagonal packings.

Full-wave simulations of a mascroscopic  random colloid metamaterial are not practical due to the   large  size of the  simulation domain which must be large enough to accommodate the entire structure. Therefore, we implement an approach based on  the supercell approximation~\cite{fan_theoretical_1995,zhao_characteristics_2008,romero-garcia_evanescent_2010}. The idea of the approach is depicted in Fig.~\ref{supercell}. The electromagnetic response of an \textit{infinite random} metamaterial is approximated with that of an \textit{infinite periodic} metamaterial whose  supercell is enough large to include many \textit{randomly} oriented meta-atoms. Consider first a periodic metamaterial with a supercell of dimensions $L \times \Delta_y \times \Delta_z$ (along the $x$, $y$, and $z$ axes, respectively) shown in Fig.~\ref{supercell1}. The different colors in the figure are employed only for visual clarity of periodicity. 
 The meta-atoms are distributed within the supercell with completely random positions and orientations without touching one another.  
The supercell is periodically repeated in the $y$ and $z$ directions. Due to the periodicity, this metamaterial slab of thickness~$L$ will diffract some part of normally incident  light into several diffraction orders.  
According to the supercell approximation, the amount of diffracted light in the periodic metamaterial provides an approximation of the incoherent scattering loss in   a corresponding random metamaterial (of the same size and with the same meta-atom concentration)  which is shown in Fig.~\ref{supercell3}. When the size of the supercell $\Delta_z$  increases (see Fig.~\ref{supercell2}),  the   number of diffraction orders also increases, and the approximation becomes more accurate. 
For our simulations, we choose  $\Delta_z (N_{\rm V}) > \lambda $ such that there are two diffraction harmonics in reflection and two diffraction harmonics in transmission. The periodicity along the $y$ direction is chosen to be sub-wavelength to reduce the overall simulation domain. Thus, the scattering analysed in this way occurs only in the plane of incidence. 

\begin{figure}[tb]
\centering
   \includegraphics[width=0.46\columnwidth]{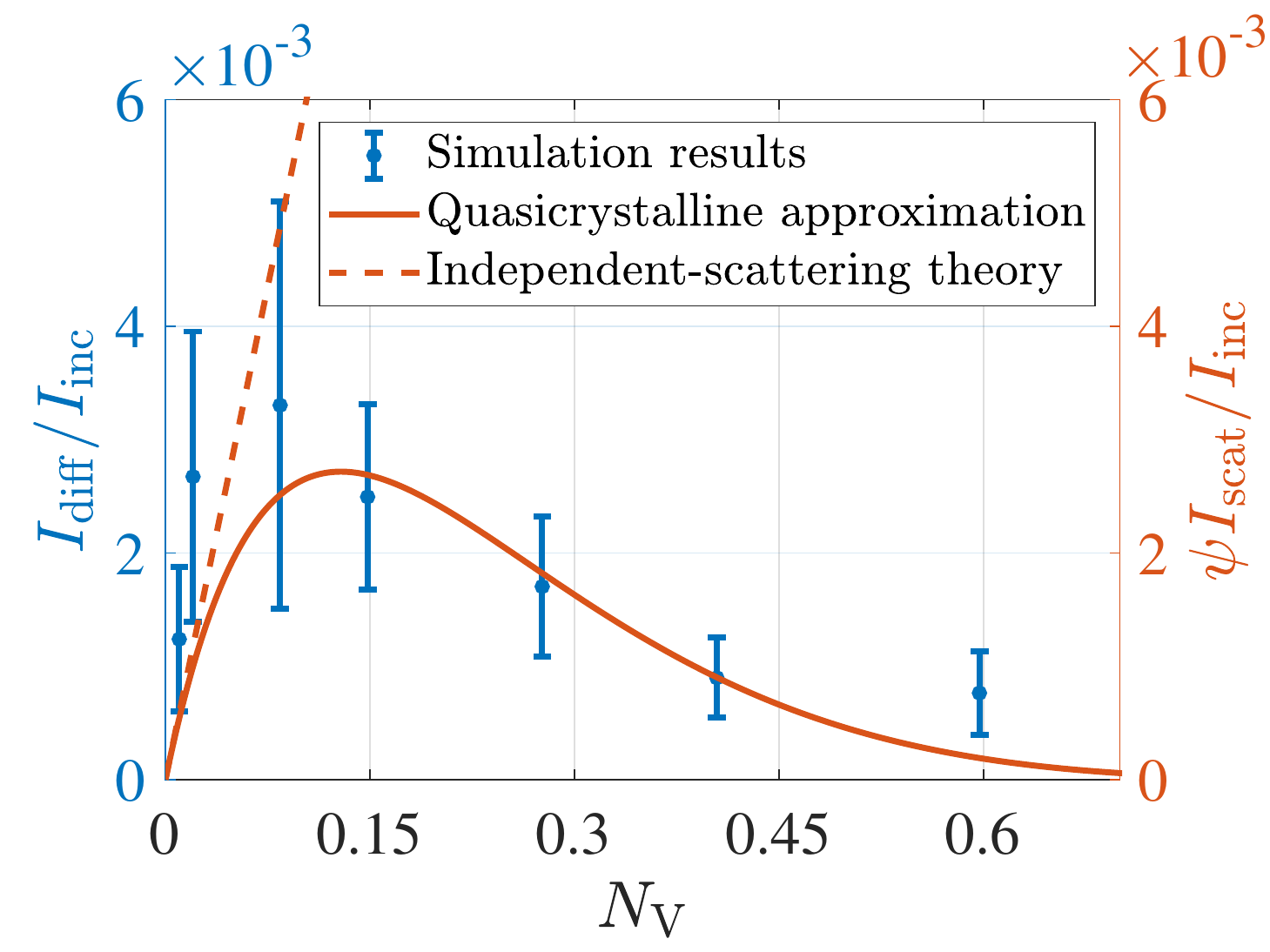}
\caption{ Comparison of the   diffraction scattering loss  in periodic metamaterial shown in Fig.~\ref{supercell1} and  the incoherent  scattering loss calculated using analytical fitting models (based on independent-scattering theory and QCA theory) for a random metamaterial shown in Fig.~\ref{supercell3}. 
The vertical blue bars  denote the standard deviation of the   values obtained in electromagnetic simulations of multiple supercell configurations.  }
\label{diff_scatt}
\end{figure}

Next, according to the supercell approximation, we analyze the attenuation coefficient~$\beta$ in a random  metamaterial  by  calculating diffraction scattering intensity for light propagating through the corresponding periodic metamaterial  of the same concentration. In what follows, we choose the periodical configuration with four diffraction orders shown in Fig.~\ref{supercell1}.  The supercells dimensions were chosen as $\Delta_y=(V_{\rm circ}/N_{\rm V})^{1/3}$, $L=3 \Delta_y$, and $\Delta_z=p \Delta_y$, where $p$ is the smallest integer for which the periodic metamaterial supports four diffraction orders. For each of the seven discrete  values of $N_{\rm V}$, we performed 15 scattering simulations of the  metamaterial supercell. In each of the 15 simulations, the supercell had the same dimensions but different random distributions and orientations of the chiral meta-atoms.  The simulated total diffracted intensity~$I_{\rm diff}$ normalized by the incident intensity~$I_{\rm inc}$ is plotted in Fig.~\ref{diff_scatt} versus the volume fraction~$N_{\rm V}$. 
Furthermore, we additionally plot in the same Figure   the incoherent scattering intensity loss calculated based on two analytical fitting models for a truly random metamaterial. The first model is based on the aforementioned QCA theory and implies $I_{\rm scat}/I_{\rm inc}=[I(0)-I(L)]/I(0)=1-e^{-\beta_{\rm corr} L} \approx \beta_{\rm corr} L$. According to the supercell approximation,   the ratios $I_{\rm scat}/I_{\rm inc}$ and $I_{\rm diff}/I_{\rm inc}$ should be proportional to one another with some unknown proportionality coefficient~$\psi$.  Substituting   $\beta_{\rm corr}$ from~\r{corr}, we chose $\psi \sigma L=6.67\cdot 10^{-23}~{\rm m}^3$ which ensures the best fitting for the simulated curve in Fig.~\ref{diff_scatt}.  As is seen, the simulated diffraction loss in the periodic metamaterial follows the same dependence on the volume concentration $N_{\rm V}$ as the theoretical orange fitting curve predicted by the QCA theory for the random metamaterial. The corresponding incoherent scattering loss for the random metamaterial predicted by the independent-scattering theory is plotted by the orange dashed line. Summarizing, one can see that incoherent scattering in the designed colloid metamaterial is in fact  suppressed by the factor $1/W(N_{\rm V})$ in accordance with the QCA theory. Thus, for high concentrations of the meta-atoms in the colloid metamaterial, the scattering loss drastically decreases when the volume concentration increases. 

\subsection{Giant isotropic chirality}
\begin{figure*}[t]
\centering
	\centering
	\subfigure[]{\includegraphics[width=0.31\linewidth]{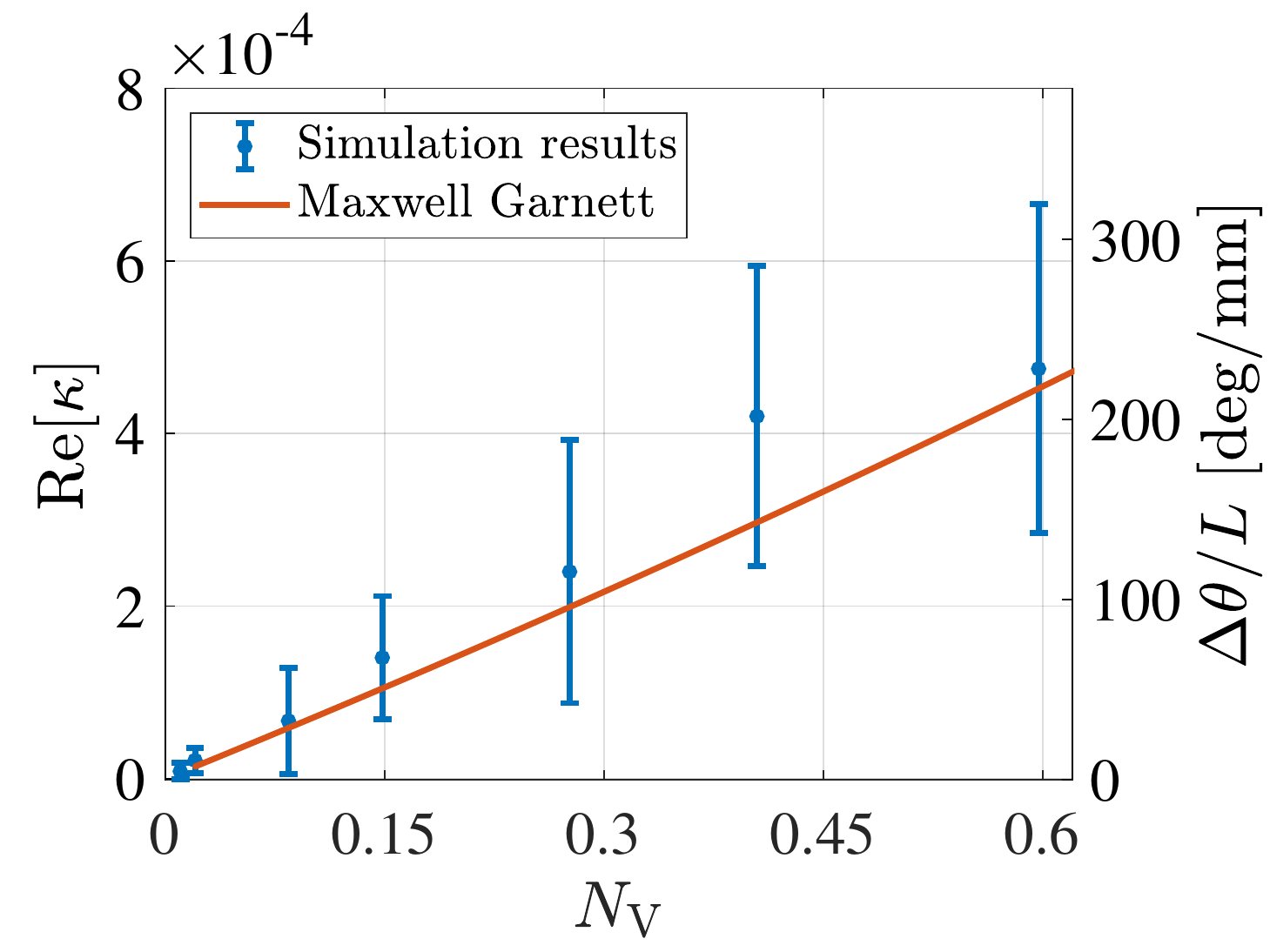} \label{chirality}}  \quad
	\subfigure[]{\includegraphics[width=0.31\linewidth]{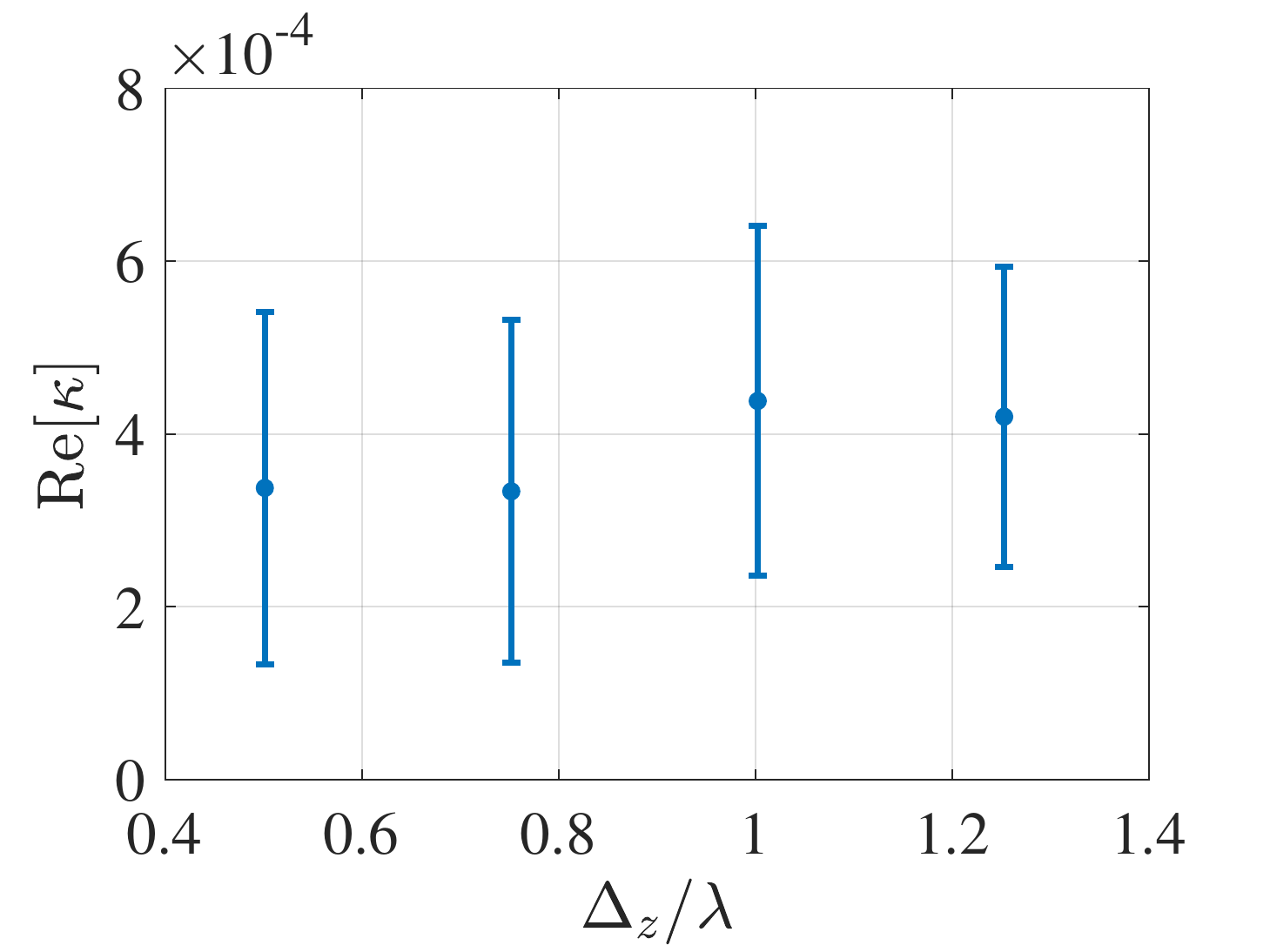}\label{chirality_vs_period}} \quad
	\subfigure[]{\includegraphics[width=0.31\linewidth]{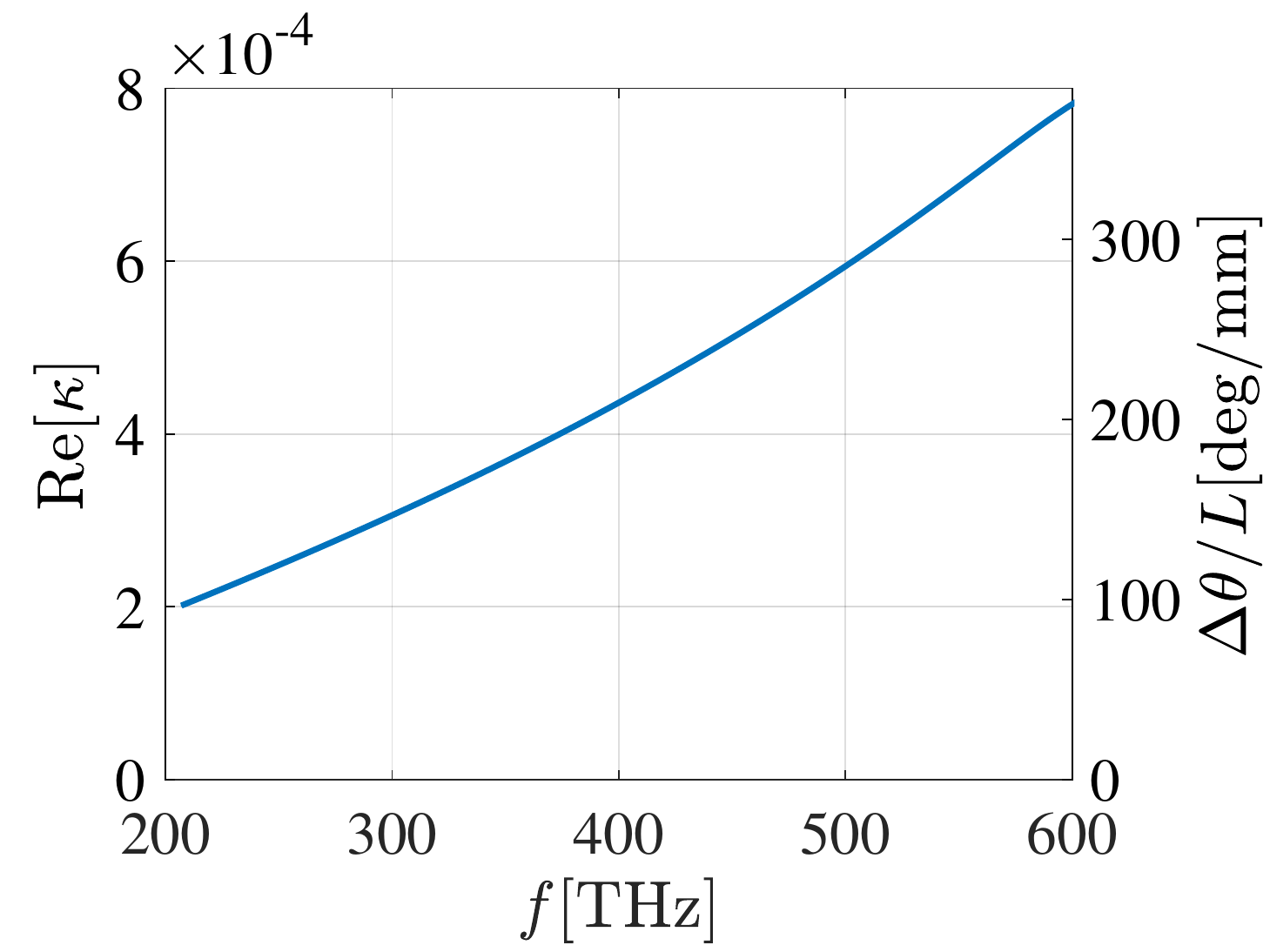}\label{dispersion}}
\caption{ (a)  The real part of the intrinsic (isotropic) chirality parameter and specific  rotation of the  metamaterial   versus the volume fraction. The plotted data corresponds to the operational frequency $f=400$~THz.  (b) The chirality parameter calculated for four different supercell sizes. In all the cases the volume fraction was fixed to  $N_V=0.4$. (c) Frequency dispersion of the  isotropic chirality parameter  and specific optical rotation of the random  colloid  metamaterial with the volume  fraction of meta-atoms~$N_V=0.6$.   }
\label{fig3}
\end{figure*}

Next, we analyze the chirality parameter versus the volume fractions for random metamaterial   shown in Fig.~\ref{supercell1}. First, the transmission and reflection data from the metamaterial slab obtained by the full-wave simulations were extracted. Using these data  and the theoretical formulation for wave propagation through a chiral slab~\cite[p.~83]{lindell_electromagnetic_1994} (including multiple reflections at the interfaces), the effective chirality parameter was calculated.
For each of the seven discrete  values of $N_{\rm V}$, we performed 50  simulations of random  metamaterial supercells.     Figure~\ref{chirality} shows the calculated mean values and the standard deviations of the chirality parameter and the specific rotation of the colloid (calculated as $\Delta \theta /L= \Re [\kappa] \omega/c$). Only the real part of parameter $\kappa$ is plotted since the imaginary part is approximately ten times smaller   due to very low absorption loss in silicon at the considered wavelengths. Therefore, the circular dichroism $\eta=\Im [\kappa] \omega L/c$ is much smaller than the optical rotation $\Delta \theta$. 
Additionally, the figure depicts the analytically calculated isotropic chirality parameter based on the polarizabilities of individual meta-atoms (see Table~\r{tab1}) and the Maxwell-Garnett mixing rule~\cite[\textsection~7.2.2]{serdyukov_electromagnetics_2001}. The theoretical curve is in good agreement with the statistical data from the full-wave simulations.  
In sharp contrast to the scattering loss, the chirality parameter  monotonically increases with $N_{\rm V}$.

\begin{table*}[tb]
   \centering
   \sffamily
\begin{tabularx}{\textwidth}{ YYYYYYY }  
     \rowcolor{black!75}
  \head{$ \lambda $ $[{\rm nm}]$} 
  & \head{$ N  $ ${[\rm m}^{-3}]$} 
  & \head{$ |{\rm tr} \{ \overline{\overline{\alpha}}_{\rm em} \}|/3 $ ${[\rm m}^3]$} 
  & \head{$|\kappa|$} 
  & \head{$ \Im(n_+ + n_-)/2 $} 
  & \head{${\rm FOM}$} 
  & \head{${\rm Ref.}$} 
  \\
      \rowcolor{black!15}    $ 750 $ 
    & $6.0\cdot 10^{14} $ 
    & $1.7\cdot 10^{-22} $ 
    & $1.0\cdot 10^{-7} $
    & $9.5\cdot 10^{-6} $
    & $0.01 $
    & \cite{mcpeak_complex_2014}
     \\
       \rowcolor{black!15}   $ 549 $ 
    & $2.4\cdot 10^{17} $ 
    & $6.3\cdot 10^{-26} $ 
    & $1.5\cdot 10^{-8} $
    & $1.5\cdot 10^{-6} $
    & $0.01 $
    & \cite{kuzyk_dna-based_2012}
     \\
        \rowcolor{black!15}  $ 560 $ 
    & $1.8\cdot 10^{17} $ 
    & $5.2\cdot 10^{-27} $ 
    & $9.3\cdot 10^{-10} $
    & --
    & --
    & \cite{shen_three-dimensional_2013}
     \\
        \rowcolor{black!15}  $ 750 $ 
    & $5.2\cdot 10^{20} $ 
    & $7.9\cdot 10^{-25} $ 
    & $4.3\cdot 10^{-4} $
    & $4.1\cdot 10^{-5} $
    & $11.3 $
    & Our work
     \\
  \end{tabularx}
  \caption{  Comparison of optical characteristics of different chiral colloids.  }
  \l{tab2}
\end{table*}


Next, it is important to examine whether  the periodicity of the metamaterial contributes to the magnitude of the chirality parameter. 
Figure~\ref{chirality_vs_period} plots the real part of parameter $\kappa$ for four different supercell sizes $\Delta_z$. The volumetric concentration of the meta-atoms in all four cases was fixed to $N_{\rm V}=0.4$.  As is seen, there is no strong correlation between the chirality parameter and the supercell size. 
Thus, we deduce that  the strong chirality is due to the  optimized meta-atoms and not due to their periodical arrangement.

Figure~\ref{dispersion} depicts the frequency dispersion of the  isotropic chirality parameter in the     metamaterial     with the   volume  fraction of meta-atoms~$N_V=0.6$.   The dispersion of silicon was used from the experimental data in~\cite{schinke_uncertainty_2015}. 
As is seen, in the frequency range from 200~THz to 600~THz, the  colloid metamaterial is nonresonant, providing high isotropic chirality  with very large bandwidth. At the higher frequencies, the meta-atoms sizes  become comparable to the wavelength in water, and the colloid cannot be considered as a homogeneous metamaterial.  
Due to the low frequency dispersion and the dielectric nature of the meta-atoms, we expect the metamaterial to have relatively robust optical response 
in the presence of polydispersity of the silicon spheres  or their imperfect assembly.

 Let us  compare the obtained chirality parameter with the previously reported values for random colloids suitable for large-scale fabrication. 
Table~\r{tab2} presents such comparison for different optical characteristics.
As is seen, the dielectric colloid metamaterial  presented in this work exhibits three orders of magnitude increase for the chirality parameter and figure of merit (expressed as ${\rm FOM}=|\kappa|/\Im(n)$) compared to previous publications. Such striking improvement is due to the two factors: The absence of plasmonic losses and scattering loss reduction due to the high-density  colloidal arrangement of the dielectric meta-atoms. 
Interestingly, analogous three orders of magnitude difference is observed between  magneto-optical  FOMs  of yttrium iron garnet and transition metal ferromagnets~\cite[\textsection~5.6.5]{coey_magnetism_2010}. Due to the higher  magneto-optical FOM (lower insertion loss for the given isolation ratio), garnets rather than ferromagnets are usually employed as Faraday-rotation components in optical isolators.  Likewise, chiral dielectric colloid proposed in this Letter has the advantage of low attenuation compared to  conventional plasmonic colloids, which makes it a better candidate for multiple applications. 
It is also worth mentioning that the designed isotropic colloid metamaterial exhibits specific rotation more than two order of magnitude larger than that of natural substances like sugar~\cite{lide_crc_2004}, and even one order of magnitude larger than that in anisotropic alpha-quartz crystal~\cite{kaminsky_experimental_2000}.

\begin{figure}[tb]
	\centering
	\subfigure[]{\includegraphics[width=0.4\linewidth]{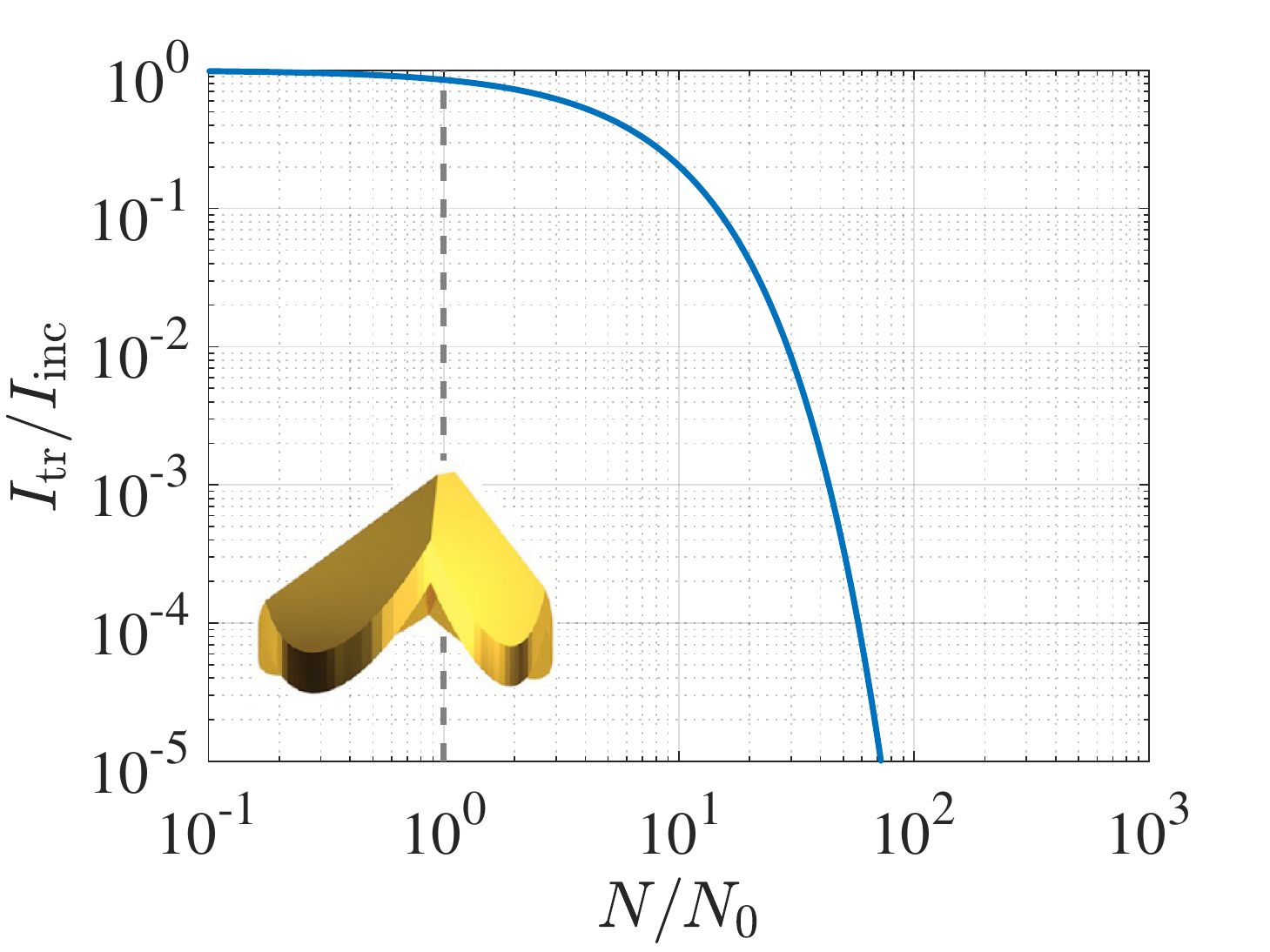} \label{fig5a}}  \quad
	\subfigure[]{\includegraphics[width=0.4\linewidth]{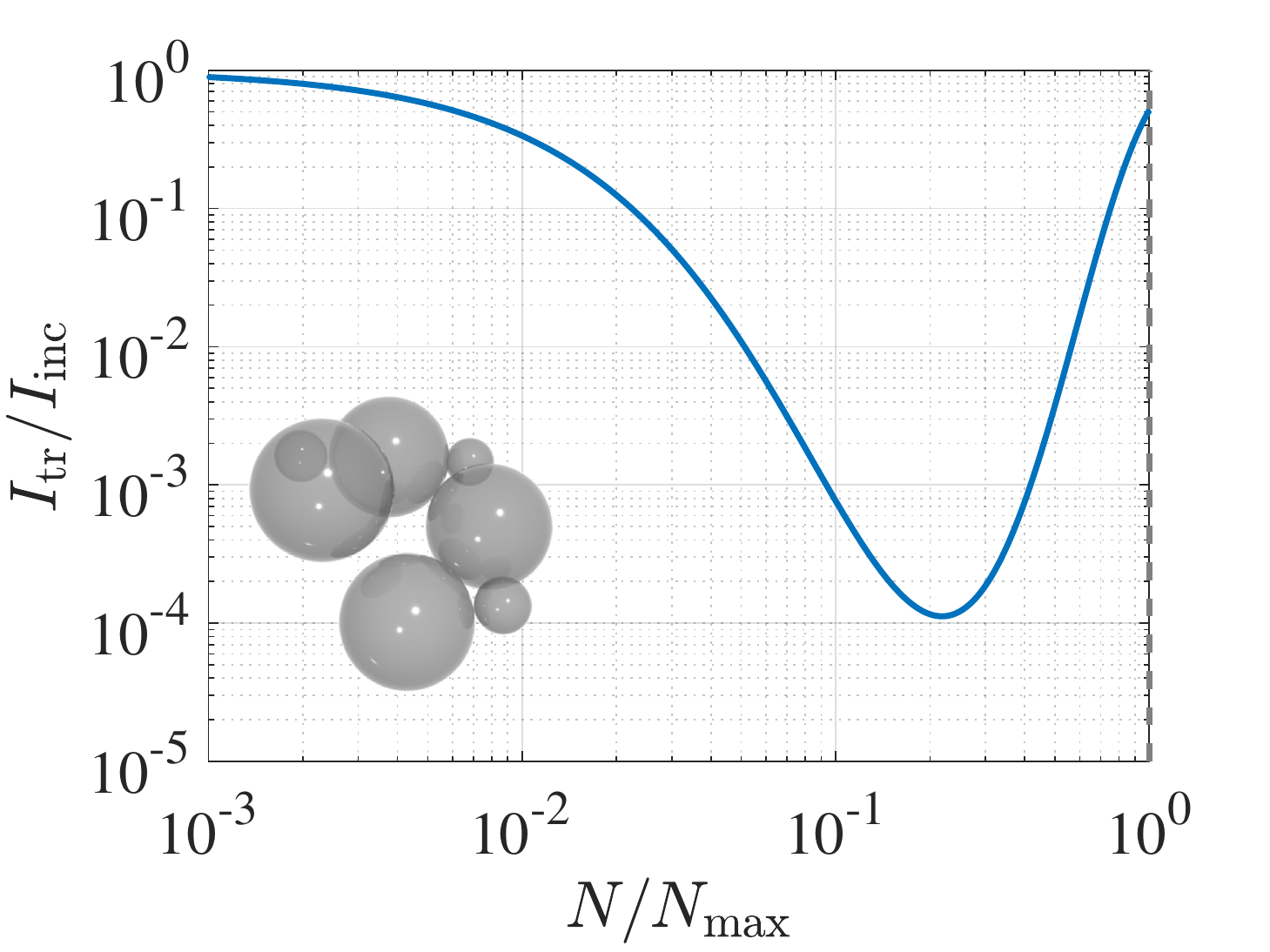}\label{fig5b}}
\caption{(a) Light transmittance through a plasmonic chiral colloid calculated based on the parameters extracted from Ref.~\cite{mcpeak_complex_2014}. The dashed vertical line denotes the initial volume concentration $N_0$. When the concentration of the plasmonic colloid increases by two orders of magnitude, the transmittance drops by more than six orders of magnitude. (b) Light transmittance through a silicon chiral colloid metamaterial proposed in this Letter. In both plots the thickness of the colloid slabs is taken $L=1$~mm. The insets depict the corresponding colloid constituents. 
 }
\label{fig5}
\end{figure}

 One can notice that the isotropic polarizability $ |{\rm tr} \{ \overline{\overline{\alpha}}_{\rm em} \}|/3 $ of the individual meta-atoms proposed in this work is several orders of magnitude smaller than that from Ref.~\cite{mcpeak_complex_2014}. This is due to the fact that in~\cite{mcpeak_complex_2014} the meta-atoms  exhibit plasmonic  resonances. In contrast, our metamaterial has nearly six orders of magnitude higher volume concentration of the meta-atoms, which results in overall increase of the chirality parameter over that in the plasmonic colloid. At  first glance it may seem that by increasing the volume concentration~$N$ in the plasmonic colloid, one can achieve stronger chirality effect. However, doing so, while enhancing  the chirality,   also inevitably  increases the absorption loss. The figure of merit  introduced  above takes into account this trade-off. Let us show that FOM does not depend on the volume concentration of the meta-atoms  in sparse colloids like in~\cite{kuzyk_dna-based_2012,shen_three-dimensional_2013,mcpeak_complex_2014}.  Since in this low-concentration case, chirality parameter is given by~\r{trace1} and the imaginary part of the refractive index $\Im (n)=N \,\Im (\alpha_{\rm eff})$ where $\alpha_{\rm eff}=\sqrt{ {\rm tr} \{ \=\alpha_{\rm ee}  \} {\rm tr} \{ \=\alpha_{\rm mm}  \} } /3$, the figure of merit becomes
\e 
{\rm FOM} = \frac{ {\rm tr} \{ \=\alpha_{\rm em}  \} }{ 3\, \Im (\alpha_{\rm eff})}.
\l{fomnew}
\f
Indeed, it depends only on the electrodynamic properties of the individual meta-atoms and not on the volume concentration  of the colloid~$N$. To further demonstrate that the giant chirality effect reported in this work cannot be achieved with conventional  plasmonic  solutions, we take as an example colloid from~\cite{mcpeak_complex_2014} and calculate the transmittance  of light $I_{\rm tr}/I_{\rm inc}$ propagating through it (reflections at interfaces are assumed negligible) for different volume concentrations~$N$:
\e 
I_{\rm tr} (N) = I_{\rm inc} \, {\rm exp} (-2k_0 L \,\Im [ n(N)]).
\l{atten}
\f
Here $ \displaystyle \Im [ n(N)]= \frac{N}{N_0} \Im [ n(N_0)] $ and parameters  $N_0=6 \cdot 10^{14}$~${\rm m}^{-3}$ and $\Im [ n(N_0)]=9.5 \cdot 10^{-6}$ are    listed in the first row of Table~\r{tab2}. Assuming $\lambda=750$~nm and $L=1$~mm, we obtain the curve shown in Fig.~\ref{fig5a}. One can see that when the concentration of the plasmonic colloid is increased by merely two orders of magnitude, the transmittance drops by more than six orders of magnitude. Therefore,  the colloid becomes impenetrable at high volume concentrations.
Note that  expression~\r{atten} does not explicitly include   packing factor  $W(N_{\rm V})$  as in~\r{corr}, and $I_{\rm tr}$ depends  linearly on concentration~$N$. This assumption is valid in the range of small volume fraction~$N_{\rm V}$ ($N_{\rm V}$ does not exceed $2.5\cdot 10^{-3}$ in the  range plotted in Fig.~\ref{fig5a}, assuming that meta-atom volume is $V_0 \approx 4.2\cdot 10^{-21}~{\rm m}^3$~\cite{mcpeak_complex_2014}).

For comparison, Fig.~\ref{fig5b} depicts the light transmittance versus the volume concentration for the silicon colloid proposed in this Letter. Since in this case the concentration is high (the maximum concentration $N_{\rm max}= 5.2\cdot 10^{20}~{\rm m}^{-3}$ is shown in Table~\r{tab2}), the transmittance intensity is now a function of  the packing factor:
\e 
I_{\rm tr} (N) = I_{\rm inc} \, {\rm exp} (-2k_0 L \,\Im [ n(N)] W(N_{\rm V}) ).
\l{atten22}
\f
Here we assumed, likewise, that $\lambda=750$~nm and $L=1$~mm.   As is seen from    Fig.~\ref{fig5b}, the attenuation does  not   monotonically  increase with increasing the volume concentration but, on contrary, it starts decaying when $N>0.22 N_{\rm max}$. This is a consequence of the scattering suppression in the quasicrystalline regime. 
 Note that for this metamaterial, the chirality parameter is monotonically increasing with $N_{\rm V}$, as seen in Fig.~\ref{chirality}. Therefore, the silicon metamaterial provides several orders of magnitude chirality enhancement compared to the plasmonic colloids, assuming the same level of light attenuation.

\section{Conclusion}
In summary, we have designed a dielectric metamaterial consisting of a random colloid of specifically designed  silicon nanoparticles  which exhibits record-high isotropic chirality with significantly suppressed   incoherent scattering. 
The suggested geometry of the chiral constituent meta-atoms favors  large-scale mass production. This is in strong contrast with the previous designs of dielectric chiral metamaterials and colloids which either suffered from low writing speeds~\cite{thiel_three-dimensional_2009} or were suitable only for low colloid concentrations (such colloids were achieved by dispersing nanoparticles from a two-dimensional area into a bulk solution)~\cite{singh_large_2018,lin_all-optical_2019,verre_metasurfaces_2017}. 
The chirality can be further increased by optimizing the geometry of the meta-atoms using, for example, via inverse design~\cite{molesky_inverse_2018}. 
It should be mentioned that although in this work, we concentrated on sub-wavelength dielectric meta-atoms, it is possible to extend this study towards meta-atoms of larger sizes. In this case, the regime of strong Mie resonances can be achieved, which would further increase the isotropic chirality of the colloid  by one or several orders of magnitude. The higher order multipole moments in isotropic samples do not contribute  to chirality due to orientational averaging~
\cite[p.~142, 227]{barron_molecular_2004}, and therefore, the theory presented in this Letter can be still  applied.  Moreover, the QCA analysis demonstrated here can be extended to the case of  meta-atoms with higher order multipole moments (see e.g.~\cite{tsang_dense_2000}).

\section*{Supporting Information}
This Section contains a simple extraction technique of the trace of the  magnetoelectric polarizability $\overline{\overline{\alpha}}_{\rm em}$ for an arbitrary meta-atom. Subsequently, this trace can be used for calculation of the isotropic chirality of the metamaterial, as shown in Eq.~(2) in the main text. 

Consider an arbitrary meta-atom in free space.
\begin{figure}[tbh]
\centering
   \includegraphics[width=0.46\columnwidth]{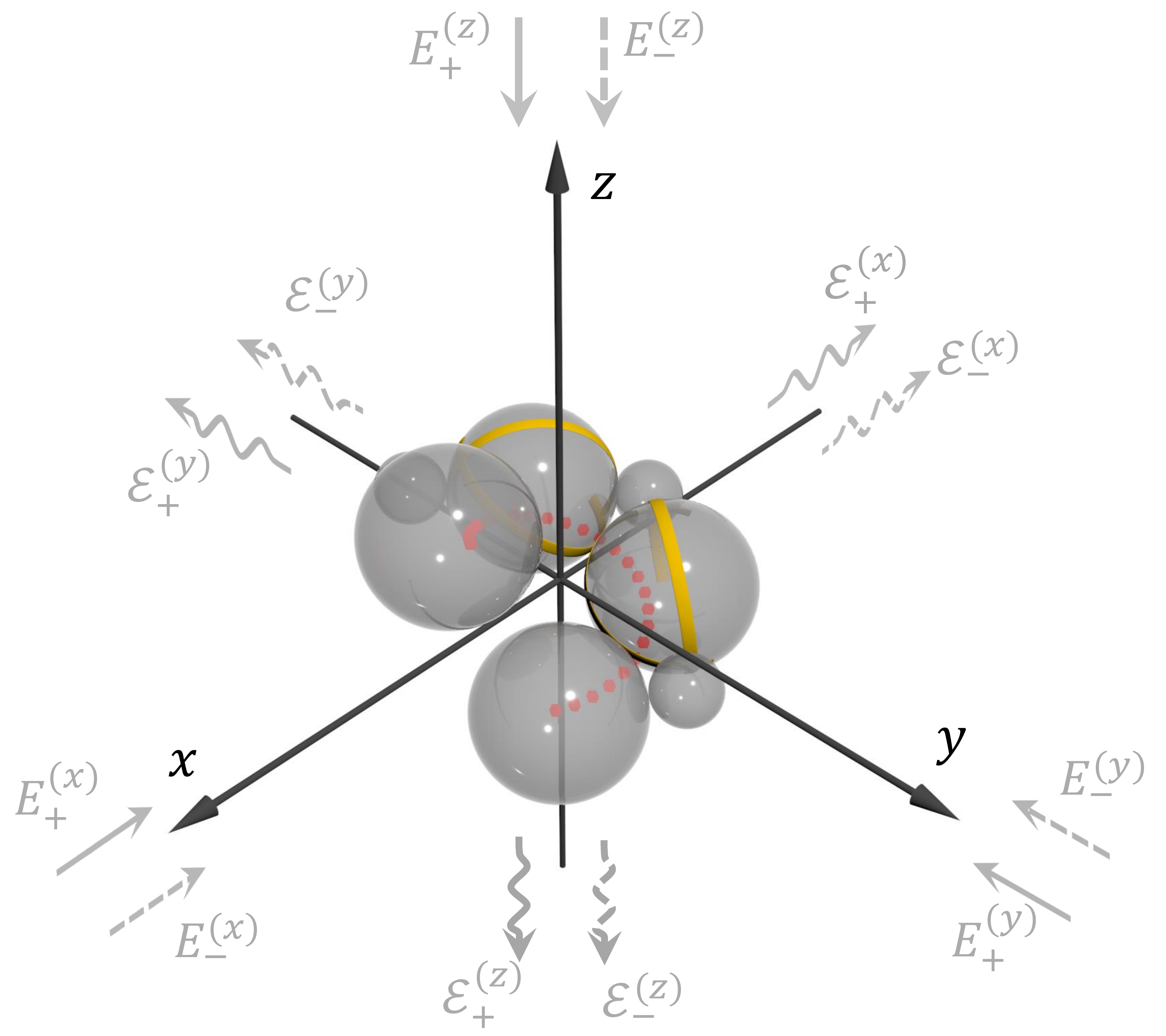}
\caption{  Designer dielectric chiral  meta-atom assembled as a cluster of crystalline silicon nanospheres. The nanospheres are DNA-coated and hybridized by DNA origami belts (shown as yellow strips) to  one another, forming a three-dimensional helical shape (shown by red dots). The straight  and curvy arrows denote the directions of the incident and forward-scattered light, respectively. The plus and minus subscripts stand for RCP and LCP light, respectively. Six measurements of the forward-scattered light with two polarizations and along three illumination directions enable  accurate and unambiguous calculation of the intrinsic (isotropic) chiral properties of the meta-atom.  }
\label{fig1}
\end{figure}
First, we illuminate it with RCP or LCP   monochromatic light with electric field amplitude $\_E_\pm= (\mp i; 0; 1)^T E_0$ and wave vector $\_k= - \omega/c \,\hat{\_y}$, as shown with orange arrows in Fig.~\ref{fig1}. The upper and bottom signs correspond to the right and left circular polarizations, respectively. The  incident magnetic field is $\_H_\pm= \_k \times \_E_\pm / (\omega \mu_0) $. Within the dipolar approximation, the induced electric and magnetic dipole moments in the meta-atom  due to the incident light  read
\begin{equation} \begin{array}{c} \displaystyle
\_p_\pm = \varepsilon_0 \={\alpha}_{\rm ee}\cdot \_E_\pm +
\displaystyle
\frac{i}{c} \={\alpha}_{\rm em}\cdot \_H_\pm,
 \vspace{1mm} \\\displaystyle
\_m_\pm = \mu_0 \={\alpha}_{\rm mm}\cdot \_H_\pm +
\frac{i}{c} \={\alpha}_{\rm me}\cdot \_E_\pm,
\end{array} \l{pm}
\end{equation}
where $\={\alpha}_{\rm ee}= \={\alpha}_{\rm ee}^T$ and $\={\alpha}_{\rm mm}= \={\alpha}_{\rm mm}^T$ are the electric and magnetic polarizability tensors, and $\={\alpha}_{\rm me}= -\={\alpha}_{\rm em}^T$ for reciprocal meta-atoms~\cite[\textsection~3.3.1]{serdyukov_electromagnetics_2001}. All the polarizability tensors in  \r{pm} have units of volume. One can then find the scattered field by the meta-atom in the forward direction $\_n=- \hat{\_y}$ at a distance $r$~\cite[p.~457]{Jackson1999}:
\e 
\_E_{\rm sc \pm} = \frac{\omega^2 e^{i \omega r/c}}{4 \pi \varepsilon_0 r c^2} \left[
(\_n \times \_p_\pm) \times \_n - \frac{\_n \times \_m_\pm}{c \mu_0} \right].
\l{esc}
\f
Here $\_E_{\rm sc +} $ and $\_E_{\rm sc -} $ represent scattered fields when the meta-atom was illuminated by  RCP (first scenario) and LCP  (second scenario)   light, respectively. 
Next, expressing the right circular polarization component of the scattered field in the first scenario as $\mathcal{E}_+ = (\hat{\_z} + i \hat{\_x}) \cdot \_E_{\rm sc +}/2$ and the left circular polarization component in the second scenario as 
$\mathcal{E}_- = (\hat{\_z} - i \hat{\_x}) \cdot \_E_{\rm sc -}/2$, we obtain the linear relation for the magnetoelectric polarizability components:
\e 
\frac{2 \pi  r c^2}{\omega^2 e^{i \omega r/c}} \left(  \mathcal{E}_+^{(y)} -  \mathcal{E}_-^{(y)} \right) = (\alpha_{\rm em}^{xx} +\alpha_{\rm em}^{zz} ) E_0,
\l{five}
\f
where the superscript~$y$ was added to indicate  that the incident light in this scenario propagates along the $y$ axis in the given basis. 
Note that in this relation, the only unknown is the sum of the polarizability components (the scattered fields can be calculated from full-wave simulations). 

Repeating the same procedure for two other sets of illuminations of the meta-atom along the $-\hat{\_x}$ and  $-\hat{\_z}$ directions  (shown with grey arrows in Fig.~\ref{fig1})  and summing up the results in the form of~\r{five}, one can finally arrive at the following expression:
\e 
{\rm tr} \{ \={\alpha}_{\rm em} \} = 
\frac{ \pi  r c^2}{\omega^2 E_0 e^{i \omega r/c}} \sum_{m=x,y,z}
\left(  \mathcal{E}_+^{(m)} -  \mathcal{E}_-^{(m)} \right),
\l{trace3}
\f
which is precisely Eq.~(3) from the main text.


\medskip
\textbf{Acknowledgements} \par 
This work was supported in part by the Finnish Foundation for Technology Promotion, by the U.S. National Science Foundation Grant No. (CBET-1641069), and by the Vannevar Bush Faculty Fellowship for U. S. Department of Defense (N00014-17-1-3030).

\medskip

%
\bibliographystyle{ieeetr}
\bibliography{Reference}



\end{document}